# Coronary CTA and Quantitative Cardiac CT Perfusion (CCTP) in Coronary Artery Disease

Hao Wu,[a] PhD; Yingnan Song,[a] BS; Ammar Hoori,[a] PhD; Ananya Subramaniam,[a] BS; Juhwan Lee,[a] PhD; Justin Kim,[a] BS; Tao Hu,[a] BS; Sadeer Al-Kindi,[b] MD; Wei-Ming Huang,[c] MD; Chun-Ho Yun,[c] MD; Chung-Lieh Hung,[d] MD; Sanjay Rajagopalan,[b,e] MD; David L. Wilson,[a,f] PhD

[a] Department of Biomedical Engineering, Case Western Reserve University, Cleveland, OH, 44106, USA
[b] Harrington Heart and Vascular Institute, University Hospitals Cleveland Medical center, Cleveland, OH, 44106, USA
[c] Department of Radiology, MacKay Memorial Hospital, Taipei, Taiwan
[d] Division of Cardiology, Department of Internal Medicine, MacKay Memorial Hospital, Taipei, Taiwan
[e] School of Medicine, Case Western Reserve University, Cleveland, OH, 44106, USA
[f] Department of Radiology, Case Western Reserve University, Cleveland, OH, 44106, USA
*dlw@case.edu

## Abstract:

**Objective.** We assessed the benefit of combining stress cardiac CT perfusion (CCTP) myocardial blood flow (MBF) with coronary CT angiography (CCTA) using our innovative CCTP software.

**Background.** By combining CCTA and CCTP, one can uniquely identify a flow limiting stenosis (obstructive-lesion + low-MBF) versus MVD (no-obstructive-lesion + low-MBF.

**Methods.** We retrospectively evaluated 104 patients with suspected CAD, including 18 with diabetes, who underwent CCTA+CCTP. Whole heart and territorial MBF was assessed using our automated pipeline for CCTP analysis that included beam hardening correction; temporal scan registration; automated segmentation; fast, accurate, robust MBF estimation; and visualization. Stenosis severity was scored using the CCTA coronary-artery-disease-reporting-and-data-system (CAD-RADS), with obstructive stenosis deemed as CAD-RADS≥3.

**Results.** We established a threshold MBF (MBF=199-mL/min-100g) for normal perfusion. In patients with CAD-RADS $\geqslant$ 3, 28/37(76%) patients showed ischemia in the corresponding territory. Two patients with obstructive disease had normal perfusion, suggesting collaterals and/or a hemodynamically insignificant stenosis. Among diabetics, 10 of 18 (56%) demonstrated diffuse ischemia consistent with MVD. Among non-diabetics, only 6% had MVD. Sex-specific prevalence of MVD was 21%:24% (M:F). On a per-vessel basis (n=256), MBF showed a significant difference between territories with and without obstructive stenosis (165±61 mL/min-100g vs. 274±62 mL/min-100g, $p<0.05$). A significant and negative rank correlation ($\rho=-0.53$, $p<0.05$) between territory MBF and CAD-RADS was seen.

**Conclusion.** CCTA in conjunction with a new automated quantitative CCTP approach can augment the interpretation of CAD, enabling the distinction of ischemia due to obstructive lesions and MVD.

# Introduction

Current imaging options for coronary artery disease are expensive, incomplete and may often require multiple modalities to confirm functionally significant coronary artery disease (CAD). Invasive coronary angiography (ICA) is the clinical standard for detecting anatomical obstructive CAD. However, the correlation between angiographic stenosis and myocardial ischemia is poor leading to the wide use of pressure wire functional flow reserve (FFR). (1) The 2021 chest pain guidelines suggest documenting anatomy with a coronary CT angiography (CCTA) exam for younger patients (< 65 years old) or when less obstructive CAD is suspected. For older patients ( >65 years old) or if obstructive CAD is suspected, confirmation of ischemia with a non-invasive function test (e.g., stress echocardiography, SPECT, PET, or cardiac MRI) for additional validation is recommended (2). The shortcomings of such an approach are apparent from the frequency of ICAs that are negative for obstructive CAD. One report estimates that >60 % of patients undergoing elective ICA had no evidence of hemodynamically significant disease (3). With ≈1M ICAs in the US every year, an effective non-invasive "gatekeeper test" could reduce many unnecessary ICAs. Many patients especially those with diabetes or hypertensive heart disease have been noted to have diffuse coronary microvascular disease or dysfunction (MVD) resulting in abnormalities in flow reserve in the absence of a functionally relevant stenosis (4–6). A recent systematic review noted a significant percentage (41%) of diffuse microvascular abnormalities or MVD in patients with non-obstructive CAD, necessitating careful differentiation as this condition does not warrant an interventional approach. Female patients have been noted to have a higher prevalence of MVD compared with male patients (7). An ideal gatekeeper test would therefore allow accurate and simultaneous delineation and distinction of a functionally significant stenosis from MVD.

CCTA is a highly sensitive tool for detecting and excluding coronary artery stenosis (8–10). The use of CT-derived FFR methods (e.g., $FFR_{CT}$ and cFFR ) have improved discrimination of hemodynamically significant CAD (11,12). In parallel studies using quantitative dynamic perfusion imaging to identify hemodynamically significant CAD have performed equally well and in some studies have been more predictive of major adverse cardiovascular events (13,14). The additional advantage of a combined CCTP+CCTA approach is that this combination can be used to identify microvascular disease. The widespread application and acceptance of stress CCTP has warranted appropriate scanner availability, approaches that correct beam hardening artifact and development of automated and pragmatic tools that allow scan registration and segmentation, facilitating rapid estimation of MBF in response to stress.

In this report, we applied a highly automated software developed by our group to quantitatively analyze CCTP images. The software enables automatic beam hardening correction, temporal scan registration, myocardium and aorta segmentation, MBF computation, and polar map conversion. We further determined the utility of this combined approach of CCTA with a quantitative myocardial blood flow (MBF) stress CCTP approach in discriminating patient subsets and the influence of gender and diabetes.

## Methods

### Data acquisition

This study was approved as a retrospective study of de-identified data by the institutional review board of Mackay memorial hospital, Taipei, Taiwan. Images were acquired starting in 2013 at Mackay memorial hospital, Taipei, Taiwan, as part of clinical protocol and shared under a data use agreement. The population consisted of 148 patients with suspected CAD who underwent CCTA and stress CCTP. We excluded 44 patients based on the following criteria: 1) age<20 years, 2) coronary artery bypass grafting, 3) acute or old myocardial infarction, 4) complete left bundle branch block, and 5) inadequate datasets such as poor image quality of CCTA or insufficient CCTP analysis. Sequential CCTA and stress dynamic CCTP were both performed using a dual-source CT system (Somatom Definition Flash; Siemens Healthineer, Forchheim, Germany) with a 128-slice detector.

CCTA acquisition and analyses are described as follows. A 20-gauge intravenous line was placed in the right antecubital vein for the administration of iodinated contrast medium and drugs related to stress perfusion. For CCTA, after a non-contrast localizing scout image was obtained, a timing bolus acquisition at the level of the aortic root was performed by administering a test bolus of 15 mL (flow rate,5 mL/sec) of contrast medium (Iopamidol 370; Bracco) followed by administration of 20 mL of saline, with a dual-syringe power injector (Ulrich medical CT motion, Germany). The timing of the beginning of acquisition for coronary CTA was determined by adding 8 seconds to the time of peak contrast enhancement in the ascending aorta to adequately enhance coronary arteries. Coronary CTA acquisition was performed with dual source, 120-kVp, and 320-mA per rotation and 0.28-second gantry rotation time with 50 mL of contrast medium followed by 40 mL of saline injected by the power injector with the injection rate of 5.0 ml/s. Prospective ECG triggering was used to cover 70- 80% of the R-R interval. Images were reconstructed with a medium smooth kernel (B26), slice thickness of 0.5 mm, and an increment of 0.3 mm. The CCTA data was evaluated by two experienced readers (CHY, WMH, with 16 and 6 years of cardiac CT experience, respectively). Readers were blinded to the subject's clinical presentation and history. Any disagreement was solved by a consensus. On a per vessel basis, stenoses were classified as the expert consensus document of the Society of Cardiovascular Computed Tomography: CAD-RADS = 0, 0% luminal diameter stenosis; CAD-RADS = 1, 1-24% stenosis; CAD-RADS = 2, 25-49% stenosis; CAD-RADS = 3, 50-69% stenosis; CAD-RADS = 4, 70-99% stenosis; and CAD-RADS = 5, 100% stenosis. CAD-RADS≥3 were considered as obstructive stenosis (15).

A stress cardiac CT perfusion was performed 10 minutes after finishing the CCTA scan. Total duration of the administration of dipyridamole (0.56mg/kg) was 4 minutes. The stress dynamic CCTP was performed six minutes after the beginning of dipyridamole infusion. The dual source scan was performed for 30 seconds, starting 6 seconds after the power injector started to administer contrast medium. Scan parameters were dual source, 100-kVp, 320-mAs per rotation, and two alternating table positions were used in the prospective ECG-triggered mode with the table moving forward and backward between the two positions in "shuttle mode," with table acceleration of 300 mm/sec. As the detector width was 38-mm with a 10% image overlap at the two positions, the imaging coverage was 73 mm. Voxel size was ~0.4 mm in plane with 3 mm slice thickness. Images were acquired in every single heartbeat for a heart rate less than or equal to 63 beats per minute and every second heartbeat for a heart rate greater than 63 beats per minute. A total of 50 mL of contrast medium was injected using the same power injector followed by 60 mL of saline, both at an injection rate of 5.0 mL/s. After stress CCTP, aminophylline was given (3 mg/kg) intravenously and delivered over 2 minutes.

**Quantitative CT perfusion pipeline**

We used a highly automated prototype software to quantitatively analyze CCTP data based on our previous works (16–20). The processing pipeline is shown in Fig 1. The quantitative CT perfusion pipeline includes (i.) aorta region of interest (ROI) detection from unregistered data, (ii.) temporal scan registration, (iii.) myocardium and aorta segmentation, (iv.) automatic beam hardening correction, (v.) MBF computation on axial images, (vi.) MBF polar map conversion, and (vii.) generating American Heart Association (AHA) segment report (21). We excluded a segment from the AHA model if the number of pixels on the polar map was less than 30% of the segment area. To determine the representative absolute MBF for each territory (LAD, RCA, and LCX), we calculated the average of the two adjacent segments that had the lowest MBF within the territory. The relative MBF was calculated by normalizing each AHA segments with the territory with highest average MBF.

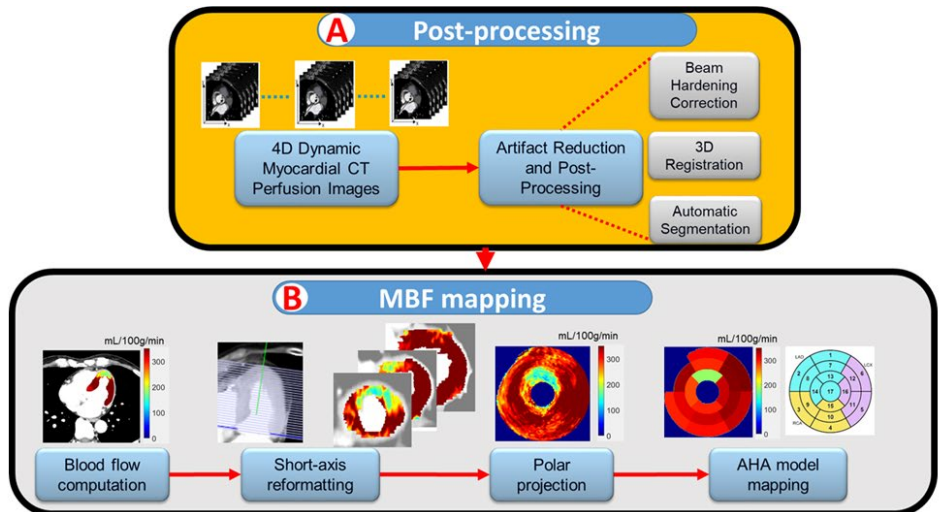

Figure 1. CCTP quantitative MBF estimation pipeline. In A, pre-processing is performed to reduce artifacts and generate the myocardial mask. In B, MBF is computed on axial data using a computational model and then transformed to short-axis data. An MBF polar map is generated. Finally, MBF is mapped to the AHA model.

**Quantitative pipeline evaluation methods**.

The pipeline was implemented using MATLAB 2022b on a computer with an Intel Core i7-8700 CPU, 3.20GHz, 64 GB RAM, and NVIDIA GTX 1080TI GPU. We trained two convolutional neural networks, one for myocardium and aorta segmentation, and one for landmarks (interventricular septum and center of left ventricle) detection in short-axis data for MBF polar map conversion, respectively. Two expert analysts manually annotated the CCTP images using AMIRA (Thermo Fisher Scientific Inc). Myocardium and aorta were segmented on the peak enhancement volume in axial images. The landmarks for polar map conversion were labeled in short-axis volume with peak enhancement. Analyst1 labeled the entire dataset, including 104 patients, to train the two convolutional neural networks. Analyst2 labeled 15 patients to perform an inter-observer study and evaluate the performance of the pipeline. The details of each step can be found in supplemental material.

We evaluated the processing pipeline's performance by comparing the territory MBF in final AHA-16 reports to those generated by a manual pipeline using the 15 patients annotated by both expert analyst1 and 2. In the manual pipeline, analysts manually performed (i.) myocardium and aorta segmentation, (ii.) short-axis reformatting, and (iii.) ventricular landmarks detection. The details of evaluation can be found in supplemental material.

**Statistical analysis**

We used the MATLAB statistic toolbox for statistical analysis (22). Continuous variables were expressed as means ± standard deviations while categorical variables were expressed as frequency and ratio (%). The unpaired Student's t-test was used for normally distributed variables to compare the means between the two groups. The Spearman's rank correlation coefficient ($\rho$) evaluated correlations between continuous MBF and ordinal CAD-RADS. To identify optimal stress MBF cutoffs on a per-territory basis, we randomly selected five vessels with CAD-RADS = 4 and five with CAD-RADS = 0. Receiver-operating characteristic (ROC) curves analysis and the Youden index were used to determine optimal stress MBF cutoffs on a per-territory basis (23). Differences were statistically significant at $P < 0.05$.

# Results

**Software validation.** We quantitatively evaluated each intermediate and final output of our software. Results for segmentation and landmark identification for mapping results to AHA segments were acceptable. In brief, for the automated segmentation module for myocardium and aorta, the average Dice score was 0.90±0.04, and 0.92±0.02, respectively. Mis-segmentation of myocardium mainly occurred around the apex (Fig. S2.) in patients with thinner myocardium near the apex area where the voxels shown partial volume appearance. In our modified AHA-16 model, we excluded the apex for MBF analysis. Euclidian distances were 2.3±1.9mm, 2.9±4.1mm, and 3.8±2.7mm, for center of LV, upper septum, and bottom septum, respectively (Fig. S3.). Small errors in the center and increased errors in the bottom of the septum were expected, as these regions coincided with errors between analysts. We also analyzed the effect of segmentation differences on results and compared results of automatic and two manual processes on territory MBF results (Figs. S4 and S5.). Agreement between analysts was excellent (slope (R) =0.99, p<0.001). The automatic method showed similarly good agreement with both analyst1 and analyst2 (slope (R)=0.98, p<0.001 and slope (R)=0.98, p<0.001, respectively). Details are shown in the Supplemental materials.

**Clinical image data analysis**. The characteristics of the 104 patients are presented in Table 1. Patient age was 58±12 years, and 41 (39%) were female. Of the 104 patients, 22 (17%) showed CAR-RADS ≥ 3 in one vessel, 13 (12%) showed CAR-RADS ≥ 3 in two vessels, and 2 (2%) showed CAR-RADS ≥ 3 in all three vessels.

Example image analyses are shown in Figs. 2. and 3. In a patient with CAD-RAD =0 (Fig. 2.), MBF was uniformly high on all views from apical to basal slices (d-f) and on Polar maps as well as modified AHA-16 displays. In Fig. 3., we show a patient with CAD-RADS = 4 with high-grade stenosis in the LAD and D1. This corresponded with reduced MBF in the LAD territory.

Table 1. Baseline characteristics of the patients.

| Parameters | All Patients (n = 104) |
|---|---|
| **Demographic parameters** | |
| Age – means± SD | 58 ± 12 |
| Female – no./total no. (%) | 41/104 ( 39 ) |
| **Cardiovascular risk factors – no./total no. (%)** | |
| Angina | 82/104 ( 78 ) |
| Hypertension | 55/104 ( 52 ) |
| Dyslipidemia | 58/104 ( 55 ) |
| Current or former smoking | 16/104 ( 15 ) |
| Diabetes | 18/104 ( 17 ) |
| **CCTA findings – no./total no. (%)** | |
| CAR-RADS < 3 | 67/104 (64) |
| 1 vessel CAR-RADS ≥ 3 | 22/104 (21) |
| 2 vessel CAR-RADS ≥ 3 | 13/104 (12) |
| 3 vessel CAR-RADS ≥ 3 | 2/104 (2) |

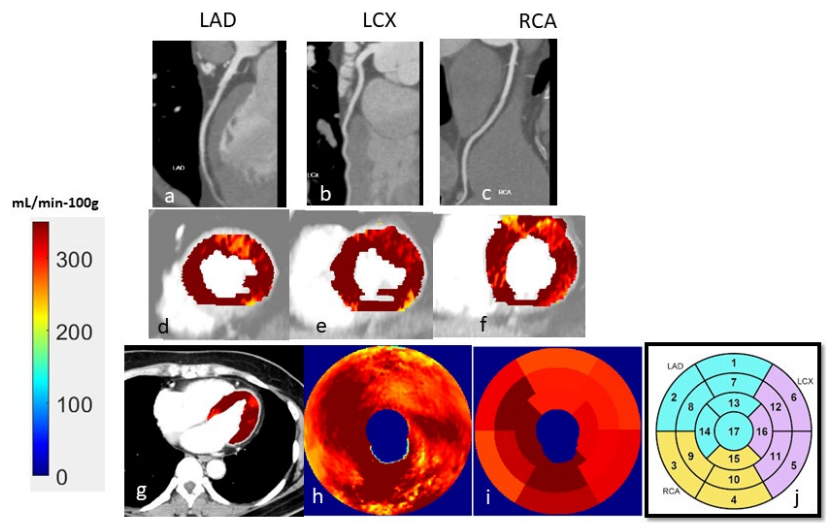

Figure 2. CCTA and CCTP analysis of a 47-year-old male patient with typical angina chest pain. CCTA (a-c) showed no luminal stenosis for all arteries. CCTP showed a normal stress MBF for all territories (d– i). The AHA sectors in (i) had a stress MBF of 320 ± 30 mL/min-100g. The map between AHA segments and the corresponding coronary artery is shown in (j).

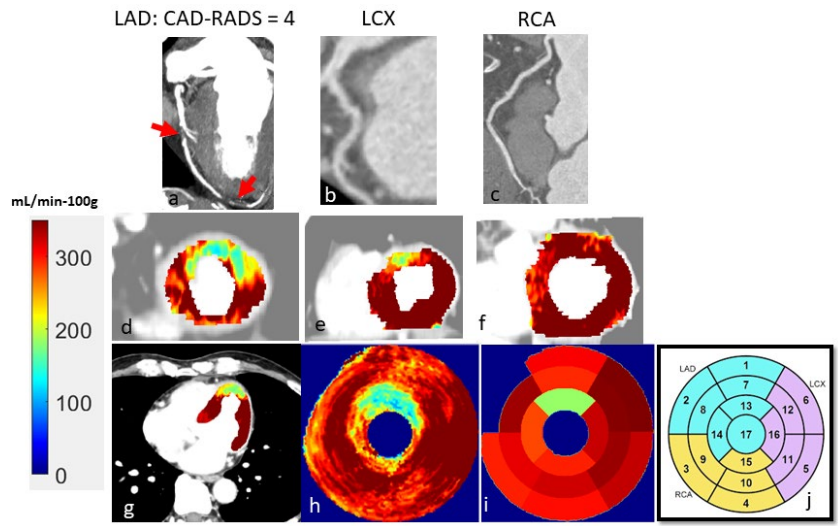

Figure 3. Male, 60-years-old patient with typical angina chest pain. CCTA (a-c) showed >70% luminal narrowing of the LAD (1-vessel disease). CCTP showed a perfusion defect with an abnormal stress MBF of 178 mL/min-100g in the territory supplied by the LAD (d– i ). The map between AHA segments and the corresponding coronary artery is show in (j).

We then determined an MBF threshold below which we deem flow to be abnormal. In order to do this, we only considered cases with unequivocal lack or presence of a significant stenosis i.e., CAD-RADS equal to 0 or 4 in non-diabetic patients to potentially avoid scenarios where significant microvascular disease in the absence of a hemodynamically significant stenosis may co-exist. We then randomly chose 5 territories for each score and performed an ROC analysis, using the presence of CAD-RADS=4 (+) (true positive) as being equivalent to abnormal flow (Fig. 4-right.). An optimal MBF threshold on the ROC was then obtained by the Youden index, giving value of 199 mL/min-100g.

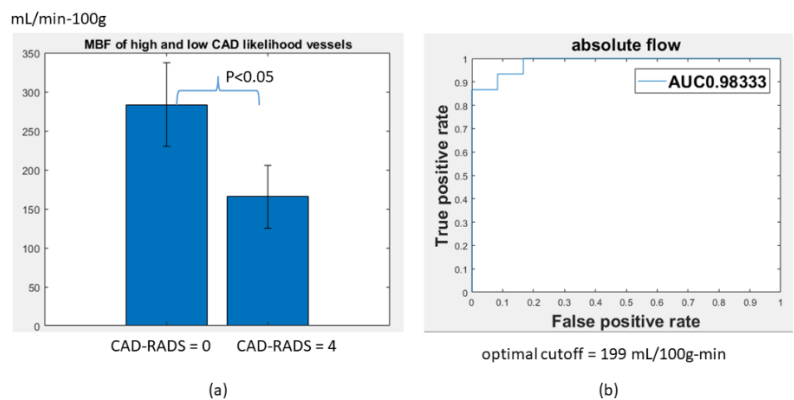

Figure 4. Obtaining an "optimal" MBF cutoff based on an analysis of unequivocal CAD-RADS cases. In (a), territory MBFs are much lower when CAD-RADS= 4 than when CAD-RADS = 0 (163 ± 37 V.S. 284 ± 58 mL/min-100g, respectively, p<0.05). In (b), assuming CAD-RADS = 4 to be (+), a ROC curve with AUC=0.98 was created by sweeping a cutoff across MBF values. The "optimal" threshold (199 mL/min-100g) for abnormal MBF was obtained by Youden index. Diabetic patients were excluded, presumably avoiding the confound of microvascular disease.

**Overall Patient Analysis.** When polar maps of all 104 patients were categorized into sub-groups, several observations were evident in a patient-centered analysis (Fig. 5.). Excluding patients with diabetes who may potentially have microvascular disease, we observed that stenosis severity showed good agreement with MBF in 50 out of 55 (90%) patients with CAD-RADS < 3 showing normal global MBF. For those with CAD-RADS ≥ 3 in one vessel, 11 out of 18 (61%) patients showed abnormal MBF territory that corresponded to the obstructed vessel. For those with CAD-RADS ≥3 in multiple vessels, 12 out of 13 (93%) patients showed abnormal MBF in at least one territory that corresponded to an obstructed vessel. Notably, for patients with CAD-RADS ≥ 3, more cases showed abnormal MBF globally in the multi-vessel group than in the 1-vessel group (10 vs. 1), and patients showed a lower MBF globally in the multi-vessel group than in the 1-vessel group (209 ± 68 mL/min-100g vs. 158 ± 47 mL/min-100g, $p<0.05$) as an internal measure of consistency.

**Non-Diabetic Sub-group.** In patients without diabetes, discordance between MBF and stenosis severity was observed in a minority of cases (Fig. 5.). In patients with CAD-RADS<3 (Fig. 5., 1st column), 5 out of 55 (9%) showed abnormal MBF with 4 out of 5 patients demonstrating abnormal MBF in the vessel in question, even though the corresponding CAD-RADS of arteries was either 1 or 2. Only 1 patient demonstrated globally abnormal MBF raising the question of diffuse microvascular disease. For non-diabetic patients with CAD-RADS ≥ 3 (Fig. 5., 3rd and 4th columns), we found that 8 out of 31 (26%) patients showed normal MBF globally. In particular, there were 4 cases with CAD-RADS = 4 that demonstrated completely normal MBF ruling out the possibility of a hemodynamically significant stenosis.

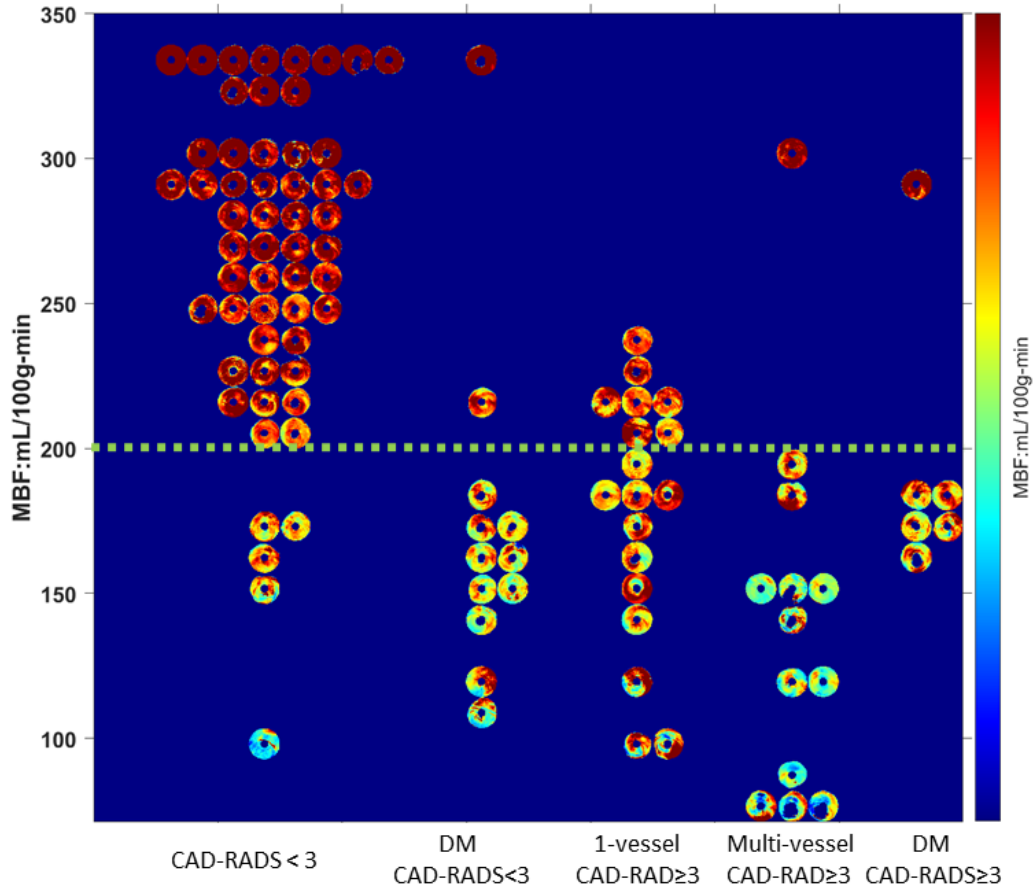

Figure 5. Polar maps of all 104 patients, where the minimum MBF in a territory gives the location along the vertical axis. All patients are divided into 5 groups based on the presence of diabetes and the number of arteries with obstructive stenosis. (a) Patients without diabetes and obstructive stenosis, CAD-RADS<3. (b) Patients with diabetes and without obstructive stenosis, DM-CAD-RADS<3. (c) Patients without diabetes and with obstructive stenosis in one vessel, 1-vessel-CAD-RADS≥3. (d) Patients without diabetes and with obstructive stenosis in multiple vessels, multi-vessel-CAD-RADS≥3. (e) Patient with diabetes and with obstructive stenosis, DM-CAD-RADS≥3. The green dashed line shows the threshold (199 mL/100g-min) for abnormal MBF. Polar maps gave results consistent with patient groupings. (See text for details.)

**Diabetic Sub-group**. Given the frequent association of abnormal MBF and coronary microvascular disease in diabetes we explored this in detail. A very high prevalence of MVD was noted in diabetic patient with CAD-RADS<3 (15 out of 18, 83%). Abnormal MBF was highly correlated to patients with diabetes, a known risk factor of microvascular disease, regardless of obstructive stenosis in coronary arteries. In Fig. 5., 15 out of 18 patients with diabetes had low MBF, often in a relatively diffuse fashion throughout the heart. Investigating sub-groups, we find that in the absence of obstructive disease (CAD-RADS<3), 10 of 12 patients with diabetes have low MBF, 1 has marginal MBF, and one has high MBF. In patients with obstructive disease (CAD-RADS ⩾ 3), 5 of 6 patients have low MBF and 1 has high MBF. We should point out that we do not know how long patients have been diagnosed with diabetes. For patients without diabetes, microvascular disease can nearly be excluded. That is, of the 55 patients without diabetes and

CAD-RADS<3, only 5 are found to have low MBF and only one of these had a severely low MBF of less than 100 mL/min-100g.

To further analyze the spatial distribution of MVD in diabetic patients, we analyzed this patient category on a per territory basis (Fig. 6.). For each patient, the distribution of MBF in each of 16 AHA segments is shown. When a distribution fell below the red dashed line threshold for abnormal MBF, ischemia was deemed present. In total, 12 out of 18 (67%) diabetic patients were deemed to have diffuse ischemia, corresponding to a low, tight distribution of MBFs. For diabetic patients without obstructive disease (green, n = 12), 8 showed diffuse ischemia in more than half of the AHA segments, and the global mean MBF was lower than the optimal cutoff. Two patients showed ischemia in one territory, and two others showed no ischemia in any territory. For diabetic patients with obstructive stenosis (purple, n = 6), four showed diffuse ischemia in multiple territories, one showed ischemia in one territory, and one showed no ischemia. Altogether, the patients with diabetes tended to have low MBF regardless of the presence of obstructive disease.

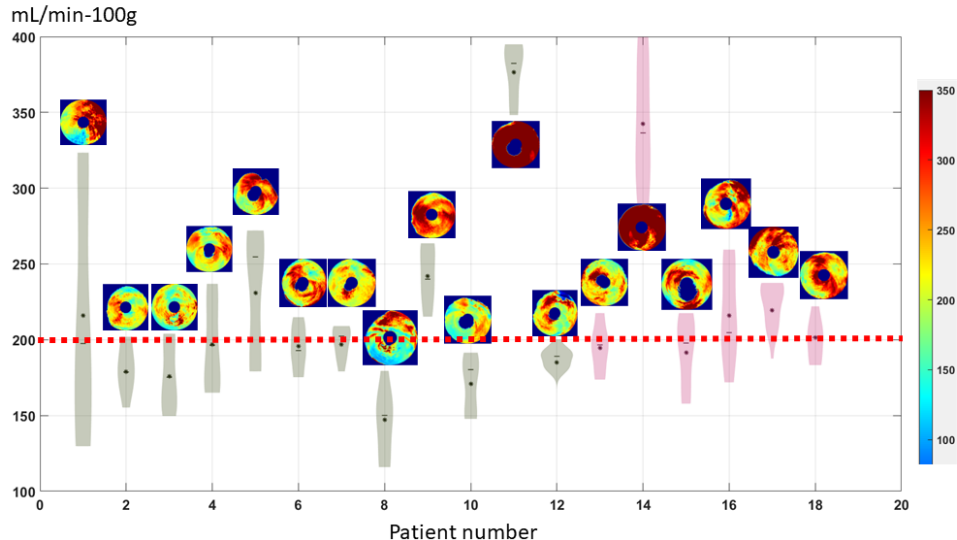

Figure 6. AHA segment distribution of diabetic patients. Diabetic patients without and with obstructive disease are in green and purple, respectively. When a distribution is below the red dashed line threshold for abnormal MBF, ischemia is deemed present. Patients 9, 11, and 14 showed normal global MBF. Patients 1, 5, and 17 showed local ischemia. The remaining 12 patients have diffuse ischemia. Note that the polar maps can be easily visually interpreted.

**Influence of Sex.** As female sex has been reported as a risk factor for microvascular disease, we analyzed gender as a variable. We defined 23 patients with low MBF and CAD-RADS<3 as having microvascular disease, regardless of the presence of diabetes. Among males and females in our study, 21% and 24%, respectively had microvascular disease. Larger numbers of patients will be required for a more thorough analysis.

**Territory analysis following exclusion of diabetic patients.** We continued with the territory analysis (Fig. 7). To determine the effect of diabetes on the analysis, we compared territories from patients with and without diabetes. Before excluding diabetic patients, mean MBFs were 259±65 mL/min-100g and 184±58 mL/min-100g in non-obstructed and obstructed territories, respectively,

p<0.001. After excluding diabetic patients, mean MBFs were 274±62 mL/min-100g and 165±61 mL/min-100g, respectively, p<0.001. By excluding diabetic patients, we reduced the confounding effect MVD on territory MBF, increased the MBF difference between non-obstructed and obstructed territories from 75 to 109 mL/min-100g, and created a clearer demarcation between territories with and without obstructive disease.

Similarly, the agreement of territory MBF and CAD-RADS stenosis severity was also improved after excluding diabetic patients (Fig. 8.). We found a significant and negative correlation between the CAD-RADS of a vessel and the absolute MBF of the corresponding territory. In Fig. 8., after excluding diabetic patients, the CAD-RADS showed a better correlation to absolute MBF than relative MBF. The MBF was the highest in CAD-RADS = 0 (286 ± 59 mL/min-100g) compared to all other categories (p < 0.05). The MBF was the lowest in CAD-RADS = 4 (156 ± 51 mL/min-100g) compared to all other categories (p < 0.05). For CAD-RADS = 3, the absolute MBF was lower than CAR-RADS = 0, 1, and 2 but only showed statistical significance with CAD-RAD = 0 and 1. The relative MBF in CAD-RADS = 4 was lower than in all other categories (p < 0.05). No statistical difference was found in CAD-RADS = 0, 1, 2, and 3. In Fig. 9., the percentage of ischemic territories (MBF< 199 mL/min-100g, blue color) increased when CAD-RADS was increased. In particular, for vessels with CAD-RADS = 3 and 4, 49% and 12% did not show downstream ischemia, respectively. For CAD-RADS = 1 and 2, there are still 21%, and 20% showed downstream ischemia, suggesting a microvascular disease.

Finally, we analyzed the percentage of territories as a function of different MBF levels (Fig. 9.), again excluding diabetic patients to limit the confounding effect of microvascular disease on MBF. As described in the figure legend, there was very good consistency between CAD-RAD scores and territory MBF.

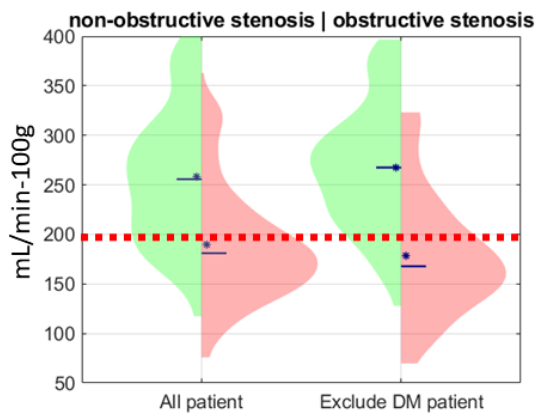

Figure 7. MBF distribution of individual territories without obstructive stenosis and with obstructive stenosis before and after excluding territories from diabetic patients. Territory MBFs are reduced in the presence of obstructive stenosis. Before excluding diabetic patients, mean MBFs are 259±65 mL/min-100g and 184±58 mL/min-100g in non-obstructed and obstructed territories, respectively, p<0.001. After excluding diabetic patients, mean MBFs are 274±62 mL/min-100g and 165±61 mL/min-100g, respectively, p<0.001. With the exclusion of diabetic patients, differences between obstructed and unobstructed territories accentuate, presumably because the microvascular disease confound has been reduced. Red dash line is the optimal MBF cutoff = 199 mL/min-100g. The blue asterisk is the median value of each distribution. The blue solid line is the mean value of each distribution.

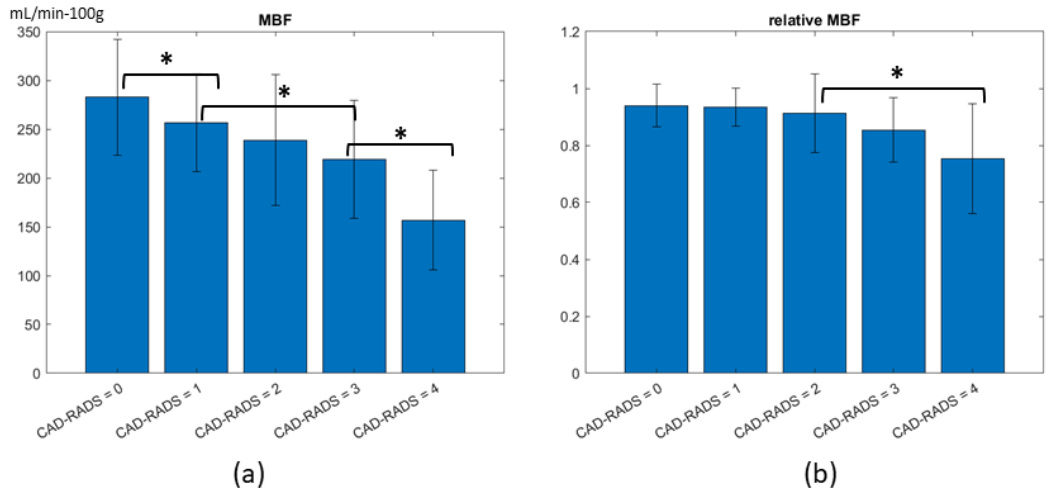

Figure 8. Association of stress MBF and CAD-RADS, territory-based analysis. (a) a moderate Spearman's rank correlation between absolute MBF and CAD-RADS (rho = - 0.53, p < 0.05). (b) a weak correlation between relative MBF and CAD-RADS (rho = -0.43, p < 0.05). *: P<0.05.

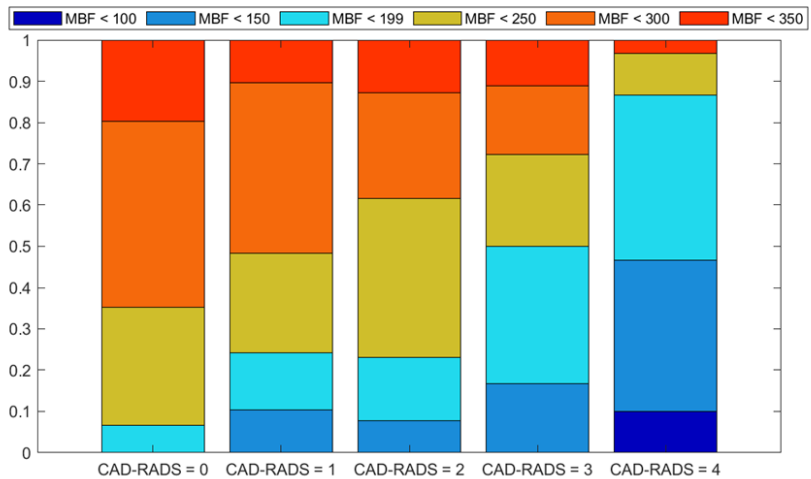

Figure 9. Percentage of individual territory MBF categorizations as a function of CAD-RAD scores for obstructive disease. Diabetic patients are excluded to reduce the confound from microvascular disease. For territories with CAD-RADS = 0, 1, and 2, 96%, 79%, and 80% of territories show normal (>199 mL/ min-100g) or greater MBF. 65%, 52%, and 38%, respectively, have MBF >250 mL/ min-100g. For these instances, no territories have MBF < 100 mL/ min-100g. For territories with CAD-RADS = 3, and 4, 49%, and 87%, respectively, have MBF <199 mL/min-100g. Percentages of high MBF values (>300 mL/ min-100g) drop precipitously as one proceeds from left to right. Altogether, the percentages falling in different groups are consistent with the extent of vascular obstruction.

## Discussion

Quantitative assessment of MBF assessment using CCTP with CCTA provides a very unique opportunity to assess the functional significance of an angiographic stenosis and concomitant abnormalities in flow reserve. These analyses could allow further characterization of stenosis related attenuation in reserve versus MVD. The main contributions of our study are: 1) The use of a prototype software for CCTP using quantitative MBF from an automated pipeline that combines beam hardening correction, temporal scan registration with automated segmentation for fast, robust quantitative MBF and visualization. 2) Delineation of a MBF threshold in patients that discriminates patients with a significant angiographic stenosis from those without. 3) Excellent agreement between the degree of stenosis severity from CCTA and absolute MBF from CCTP especially after excluding patients with diabetes who have a high likelihood of microvascular disease. 4) High prevalence of likely MVD in patients without obstructive CAD especially in those with diabetes and women indicating the presence of microvascular disease.

We further refined a highly automated pipeline for analyzing MBF in CCTP images using innovations that have been rigorously tested and reported in previous publications (16–20). To correct for beam hardening artifacts, we used an image-based automated beam hardening correction (ABHC) that we have described previously.(20) We have shown that the ABHC2 algorithm outperformed other methods in terms of HU value and final MBF estimation in digital dynamic anthropomorphic phantoms and in preclinical porcine experiments that included gold standard virtual keV images free of beam hardening. (20). We created a robust physiology-based perfusion model, which proved to be more accurate than 8 other approaches when applied to very realistic simulated data (17). To reduce the effects of noise and obtain better MBF estimates, we developed the simple linear iterative clustering algorithm with robust perfusion quantification (SLICR) method (18) applied in the current report. In the case of high noise (50% standard x-ray dose level), SLICR improved both standard deviations and means as compared to a singular value decomposition-model independent method with spatio-temporal bilateral filter (101 ± 12 vs. 54 ± 24 mL/min-100g, $p < 0.05$, actual MBF = 100 mL/min-100g). It gave MBF estimates of 101±6 mL/min-100g for an actual MBF of 100 mL/min-100g, at standard x-ray dose levels and gave estimates resistant to increases in noise. SLICR also greatly reduced computation time, making volumetric MBF measurements feasible.

Our results show that in stress CCTP, the absolute MBF is better correlated to CAD-RADs than relative MBF ($\rho$ = - 0.53 vs. $\rho$ = -0.43). Previous literatures generally support higher diagnostic accuracy by using relative MBF than absolute MBF in stress CCTP (24,25). However, Kajander et al. reported a higher diagnostic accuracy by using absolute MBF than relative MBF with positron emission tomography (PET) (26). Another PET study by Stuijfzand et al. did not find significant improvement in detecting hemodynamically significant stenosis with relative MBF compared to absolute MBF alone (27). Possible explanations include subclinical atherosclerosis in reference arteries, presence of diffuse ischemia in patients with multi-vessel disease, and the inherent noise and variations associated with ratio measurements of relative MBF. Similar issues have been reported in PET studies, which showed lower diagnostic accuracy for relative flow measurements compared to absolute MBF measurements (27–30).

Various MBF values have been reported in the literature for determination of ischemia. PET scans thought to be the most accurate for measuring blood flow use values from 185 to 250 mL/min-100g for determination of obstructive stenosis (31). In cardiovascular magnetic resonance studies (CMR), the cutoff MBF varies from 129 to 194 mL/min-100g (32,33). The optimal cut-off to discriminate stenosis of CAD-RADS = 4 in our study was 199 mL/min-100g. The SPECIFIC study measured absolute MBF from CCTP to detect obstructive CAD across multiple centers. The optimal threshold for absolute MBF to identify obstructive CAD was 142 mL/min-

100g (34). The method for computing MBF (max-upslope) contributes to the lower estimated threshold in the SPECIFIC study. Our previous work demonstrated that the max-upslope method significantly underestimated the MBF in both simulated and in-vivo animal studies (17). Similarly, the low values reported for MRI might also depend upon the computational method.

We found that absolute CCTP MBF decreased with increasing stenosis, especially in patients without diabetes (thus avoiding the confound of possible MVD). The inverse relationship between stenosis severity and downstream myocardial ischemia was reported in PET studies previously (35,36). From Di Carli et al., PET flow reserve measurement was inversely and moderately related to stenosis measured by coronary angiography (35). The flow reserve was significantly lower in stenosis between 70% and 90% than between 50% and 70%. Naya et al. found a significant but moderate correlation between the stenosis severity by CCTA and the downstream MFR measured by PET (36). Tonino et al. used invasive FFR to assess the association of hemodynamic stenosis and angiographic severity and demonstrated that intermediate stenosis does not often cause myocardium ischemia (37). In addition, the coronary collateral flow can serve as an alternative source of blood supply to the myocardium territory fed by an artery with obstructive stenosis, therefore reducing ischemia (38).

We found that in patients *without* obstructive stenosis, 15 out of 67 (22%) showed reduced MBF suggesting an opportunity for deeper analysis. In Fig. 6, for the 12 diabetic patients without obstructive stenosis (DM-CAD-RADS<3), 10 had reduced MBF (indicating MVD) in one or more territories (83%) compared 5 patients (6%) in those without diabetes. MVD Studies show that MVD is prevalent in patients with symptoms of CAD (39–41). In the ISCHEMIA trial, 13% of patients with moderate or severe ischemia confirmed by core lab did not show evidence of obstructive CAD on CCTA (39). Murthy et al. used PET to assess myocardial ischemia in 1218 symptomatic patients without obstructive CAD and showed that 51% of men and 54% of women had MVD (40). From the MBF sector distribution for patients with diabetes (Fig. 6.), 12/18 diabetic patients showed diffuse ischemia. Our finding confirmed that MVD is prevalent in diabetes, as identified by other quantitative methods (42,43). The significantly lower MBF in patients with diabetes compared to normal patient was demonstrated in PET (42). Using intracoronary thermodilution technique, Gallinoro et al. showed higher microvascular resistance in patients with diabetes as compared to those without diabetes (43).

Despite the obvious limitation of limited data from a single site, our approach paves the way to proceed with the use of quantitative MBF from CCTP. The complementary value of MBF with an assessment of obstructive disease from CCTA provides a unique comprehensive test combining anatomic and functional assessment in a single test. This helps to better identify patients who would benefit from revascularization and identifies patients who have MVD, improving precision phenotyping and personalized therapeutic approaches.